\documentclass{jpconf}
\usepackage{graphicx}
\usepackage{amssymb}
\usepackage[symbol]{footmisc}
\usepackage{aas_macros}

\newcommand{\msun}{\,\rm M_\odot}
\newcommand{\be}{\begin{equation}}
\newcommand{\ee}{\end{equation}}
\newcommand{\f}{\frac} 

\newcommand{\rtwo}{r_{\rm 200}}

\newcommand{\Vmax}{V_{\rm max}}
\newcommand{\rVmax}{r_{\rm Vmax}}

\newcommand{\dd}{{\rm d}}
\newcommand{\grad}{\vec{\nabla}}
\newcommand{\pkdgrav}{\textsc{pkdgrav}}
\newcommand{\VL}{\textit{Via Lactea}}

\begin{document}
\title{The Via Lactea INCITE Simulation: Galactic Dark Matter Substructure at High Resolution}

\author{Michael Kuhlen$^1$, J\"urg Diemand$^2$, Piero Madau$^2$, Marcel Zemp$^2$}

\address{$^1$School of Natural Science, Institute for Advanced Study, Einstein Lane, Princeton, NJ, 08540}
\address{$^2$Department of Astronomy \& Astrophysics, University of California Santa Cruz, 1156 High Street, Santa Cruz, CA, 95064}

\ead{mqk@ias.edu}

\begin{abstract}
It is a clear unique prediction of the cold dark matter paradigm of
cosmological structure formation that galaxies form hierarchically and
are embedded in massive, extended dark halos teeming with self-bound
substructure or "subhalos". The amount and spatial distribution of
subhalos around their host provide unique information and clues on the
galaxy assembly process and the nature of the dark matter. Here we
present results from the \VL\ INCITE simulation, a one billion
particle, one million cpu-hour simulation of the formation and
evolution of a Galactic dark matter halo and its substructure
population.
\end{abstract}

\section{Introduction}

According to the standard model of cosmology our Universe is made up
mostly of mysterious dark energy (70\%) and cold dark matter (CDM,
25\%). Ordinary matter (baryons) makes up only 5\% of the mass density
in the Universe today. The dark energy component is homogeneous and
structure formation is dominated by the CDM component.  In the last
few years cosmological N-body simulations have successfully uncovered
how the cold dark matter distribution evolves from almost
homogeneous initial conditions near the Big Bang into the present
highly clustered state \cite{diemand2005, springel2005}. CDM halos are
built up hierarchically in a long series of mergers of smaller halos.
Early low resolution simulations and simple analytical models found
that the end products are smooth, triaxial halos. Higher resolution
cosmological N-body simulations later modified this simplistic
picture. The merging of progenitors is not always complete: the cores
of accreted halos often survive as gravitationally bound subhalos
orbiting within a larger host system. CDM halos are not smooth, they
exhibit a wealth of substructure on all resolved mass scales
\cite{moore1998}. Subhalos are now believed to host cluster galaxies
and the satellite galaxies around the Milky Way.

In 2007 we received an Innovative and Novel Computational Impact on
Theory and Experiment (INCITE) grant for one million cpu-hours on Oak
Ridge National Laboratory's \textit{Jaguar} Cray XT3 supercomputer, to
run a one billion particle simulation of the formation and evolution
of a Milky Way scale dark matter halo in an expanding cosmological
volume. The resulting simulation, dubbed \VL\footnote[2]{Actually this
  simulation is the second in a series of simulations of
  ever-increasing resolution and is referred to in the literature as
  \textit{Via Lactea II}.}, has resolved the subhalo population at
unprecedented detail and closer to the host halo's center than ever
before. In the following we describe some of the technical details
that lie behind this computational effort, and present some of the
first scientific results that have emerged from this study.

\section{Technical Details}

Measurements of the cosmic microwave background by the Wilkinson
Microwave Anisotropy Probe (WMAP) satellite \cite{spergel2007} have
given us an exquisite view of the state of the universe a mere
$\sim$100,000 years after the big bang. At this time the universe was
very simple: a complete statistical description of the primordial
temperature fluctuations observed by WMAP requires only six
parameters. This simple statistical description of the early universe
allows us to generate an initial particle distribution at redshift
$\sim100$ (only 15 million years after the Big Bang) with small
linear density perturbations matching the WMAP measurements. Our
simulation then follows the gravitational collapse of these
fluctuations into the highly non-linear clustering regime today.

\subsection{The N-body technique}

The dark matter density field is sampled using a finite number $N$ of
Lagrangian tracer particles, whose evolution is governed by Newton's
equations of motion in coordinates co-moving with the expansion of the
universe \cite{peebles1980}:
\be
\f{\dd \vec{x}}{\dd t} = \vec{v}
\label{eq:eom1}
\ee
\be
\f{\dd \vec{v}}{\dd t} + 2 H(a) \vec{v} = -\f{1}{a^2} \grad \phi.
\label{eq:eom2}
\ee
Here $a$ is the expansion scale factor, $H(a)=\dot{a}/a$ the Hubble
constant, and $\phi$ a modified gravitational potential
\be
\phi = \Phi + \f{1}{2} a \ddot{a} x^2,
\ee
whose field equation is given by
\be
\nabla^2 \phi = 4 \pi G \rho a^2 - \Lambda a^2 + 3 a \ddot{a},
\label{eq:phi_field}
\ee
where we are allowing for the contribution of a vacuum energy of
density $\rho_v = \Lambda/(8 \pi G)$. Using the $2^{\rm nd}$ Friedman
equation
\be
\f{\ddot{a}}{a} = -\f{4}{3} \pi G \rho_b(t) + \f{\Lambda}{3}
\ee
equation~\ref{eq:phi_field} reduces to
\be
\nabla^2 \phi = 4 \pi G (\rho - \rho_b) a^2.
\ee
Because the effective gravitational potential is determined by the
density contrast $(\rho - \rho_b$), the computational error on the
gravitational potential is set by how accurately the density field is
sampled. A larger $N$ will thus result in a more faithful
representation of the dark matter dynamics, and this has been the
driving force leading to ever increasing problem sizes and the use of
powerful supercomputers for their solution.

\subsection{\pkdgrav, a hierarchical tree code}

\pkdgrav\ \cite{stadel2001}, the code that we used for the \VL\
simulation, is an example of a hierarchical tree code, in which the
particles are grouped together hierarchically and the gravitational
force on a particle from each group is approximated by a multipole
expansion up to low order. The hierarchical grouping is realized in a
tree structure, in which the whole computational domain is split up
into ever smaller sub-regions containing successively smaller numbers
of particles, down to the leaf nodes of the tree which contain only a
small number of particles. The calculation of the gravitational force
felt by each particle proceeds by ``walking'' the tree, starting with
the root node that covers the entire computational domain. A
cell-opening criterion is applied to each node to determine whether to
use its multipole approximation, ending the descent along this branch,
or if the cells needs to be ``opened'' and its daughter nodes
examined.

The use of such a hierarchical multipole expansion requires only $\log
N$ interactions per particle, much less than the $N-1$ separate force
calculations required with direct-summation, and this allows much
larger particle numbers to be used. The accuracy of the force
calculation is determined by the order to which the multipole
expansion is carried out and by how many nodes it is applied to.
Increasing the order of the expansion allows greater accuracy, but
comes at an additional computation cost. \pkdgrav\ extends the
multipole expansion up to fourth order.

\pkdgrav\ employs a spatially balanced binary tree, a relative of the
particle balanced \mbox{k-d} tree. Starting with the root node, every
node is recursively split up into two rectangular pieces of equal
volume by bisecting the tight bounding box containing all particles in
the node.  This splitting continues down to the leaf nodes, or
``buckets'', which are allowed to contain at most 16 particles. Such a
tree has been shown to be asymptotically superior to density based
decompositions like the \mbox{k-d} tree \cite{anderson1996}.

The cell opening criterion applied by \pkdgrav\ is
\be
r_{\rm open} = \f{2}{\sqrt{3}} \f{b_{\rm max}}{\theta},
\ee
where $b_{\rm max}$ is the maximum distance from any point in the cell
to the cell's center of mass, and $\theta$ is a user-specified
accuracy parameter typically set to $\theta=0.7$ for $z<2$ and
$\theta=0.5$ at higher redshifts. If during the walk a cell is deemed
sufficiently far from bucket $B_i$ to allow its multipole
approximation, then it is added to $B_i$'s list of particle-multipole
interactions. Alternatively, if a particular branch is followed all
the way down to its leaf node, then all the particles in that bucket
are added to $B_i$'s particle-particle interaction list.

In order to avoid unphysical two-body collisions the potential is
softened for small separations, $|\vec{x}_i - \vec{x}_j|<\epsilon$.
This is accomplished by using a single particle density distribution
that is given by a Dirac $\delta$-function convolved with a softening
kernel, here taken to be a compensating $K_1$ kernel
\cite{dehnen2001}. At separations greater than about $2 \epsilon$ the
potential becomes exactly Newtonian.

\subsection{A new dynamical time step criterion}
An accurate determination of the gravitational potential is a
necessary, but not sufficient, requirement for a reliable solution to
the equations of motion of dark matter. Of equal importance is the
method used to integrate the equations and the choice of time step.
Cosmological simulations typically exhibit a dynamic range of
densities spanning many orders of magnitude. Denser particle
distributions require shorter time steps for a given accuracy, but
using a global time step set by the densest regions in the simulations
becomes computationally impractical, since the majority of the
particles reside in much lower density regions. A more efficient use
of computational resources can be achieved by the use of individual
time steps that adapt to the local density. \pkdgrav\ employs an
adaptive leapfrog integrator, which allows such adjustable time steps.
The time steps are quantized in powers of two of the basic time step
$T_0$.

\begin{figure}
\begin{center}
\includegraphics[width=0.6\textwidth]{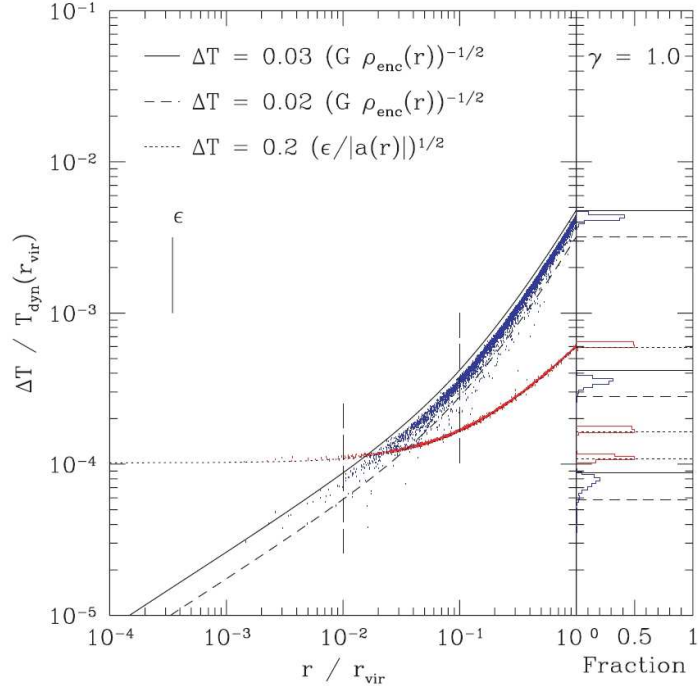}
\caption{The distribution of time steps for a density profile of slope $\gamma=1$, for the standard time step $\Delta T \propto \sqrt{\epsilon/|a(r)|}$ (in red) and for the new dynamical time step $\Delta T \propto 1/\sqrt{G \rho(<r)}$ (in blue). From Zemp et al. (2007) \cite{zemp2007}.}
\label{fig:tdyn}
\end{center}
\end{figure}

A widely used time step criterion is based on a rather ad-hoc
combination of the force softening and the acceleration a given
particle feels, $\Delta T \leq 0.2 \sqrt{\f{\epsilon}{|\vec{a}|}}$.
This criterion results in a very tight time step distribution and has
been shown \cite{power2003} to produce accurate results at all but the
very highest densities. However, this criterion is not directly
related to the local dynamical time, and it has been shown
\cite{zemp2007} that for very high dynamic ranges ($\gtrsim 10^3$) it
is not quite adaptive enough, resulting in overly conservative time
steps in the outer low density regions and insufficiently short time
steps in the very centers of dark matter halos. Instead Zemp et al.
\cite{zemp2007} advocate a time step criterion that is directly set by
the local dynamical time a particle feels,
\be
\Delta T \leq \eta \f{1}{\sqrt{G \rho_{\rm enc}(r)}},
\ee
where $\rho_{\rm enc}$ is the enclosed density from the particle's
location to the dominant dynamical structure, and $\eta$ is typically
set to $0.02-0.03$. The challenge is to quickly determine the dominant
structure governing the orbit of a given particle, and Zemp et al.
\cite{zemp2007} have achieved this by cleverly making use of the
hierarchical tree structure used for the force calculation.
Figure~\ref{fig:tdyn} (taken from \cite{zemp2007}) shows a comparison
of the time step distribution using the old and new time step
criteria, and reveals that the new time step is more adaptive, with
smaller time steps close to the center and larges one in the outer
regions. The most recent version of \pkdgrav\ contains an
implementation of this new algorithm, and it allowed our \VL\
simulation to push to unprecedented central densities at a manageable
computational cost.

An unfortunate side effect of the use of individual time steps is that
it destroys the symplectic nature of the integrator. This would be
highly undesirable for any long term integration of planetary orbits,
for example. In cosmological simulations, however, dark matter
particles typically complete a comparatively much smaller number of
orbits over a Hubble time, and so any secular error is not likely to
grow significantly during the simulation. In this case the
computational advantage of using individual and adaptive time steps
outweighs the negative side-effects.

\begin{table}[ht]
\begin{center}
\caption{Simulation Parameters\label{table:simulation}}
\begin{tabular}{cccccccccc}
\hline \hline
$L_{\rm box}$ & $\epsilon$ & $z_i$ & $N_{\rm hires}$ & $M_{\rm hires}$ & $\rtwo$ & $M_{200}$ & $\Vmax$       & $\rVmax$ & $N_{\rm sub}$ \\
(Mpc)         & (pc)       &       &                 & ($\msun$)       & (kpc)     & ($\msun$) & (km s$^{-1}$) & (kpc)    &               \\
\hline \\ [-2ex]
 40.0 & 40.0 & 104 & $1.05 \times 10^9$ & $4.10 \times 10^3$ & 402 & $1.93 \times 10^{12}$ & 201 & 60.0 & 53,653 \\
\hline
\end{tabular}
\end{center}
{\small Box size $L_{\rm box}$, (spline) softening length $\epsilon$,
  initial redshift $z_i$, number $N_{\rm hires}$ and mass $M_{\rm
    hires}$ of high resolution dark matter particles, host halo
  $\rtwo$ (radius enclosing 200 times the mean density), $M_{200}$
  (mass within $\rtwo$), $V_{\rm max}$ (the peak of the circular
  velocity curve, $v_{\rm circ}=\sqrt{GM(<r)/r}$) and $\rVmax$ (the
  radius at which $\Vmax$ occurs), and number of subhalos within
  $\rtwo$ $N_{\rm sub}$ at $z=0$. Force softening lengths $\epsilon$
  are constant in physical units back to $z=9$ and constant in
  co-moving units before. }
\end{table}

\subsection{Multi-mass initial conditions}
The \VL\ simulation makes use of multi-mass initial conditions
\cite{bertschinger2001, power2003}, which allow the halo of interest
to be resolved with very high resolution, while at the same time
allowing for the large scale cosmological tidal field to be
represented at much lower resolution. The overall computational volume
is (40 Mpc\footnote[3]{One megaparsec (Mpc) is approximately 3 million
  light years.})$^3$, but at the end of the simulation 95\% of all
high resolution particles are found within the inner (2 Mpc)$^3$. In
total the \VL\ simulation consists of a hierarchy of three particle
masses: (16,689,261, 28,645,888, 1,048,772,608) particles with masses
(32,768, 64, 1) times the high resolution particle mass of $M_{\rm
  hires}=4.10 \times 10^3 \msun$. The remaining parameters of the
\VL\ simulation are summarized in Table~\ref{table:simulation}.

\section{Results}

\begin{figure}
\begin{center}
\includegraphics[width=\textwidth]{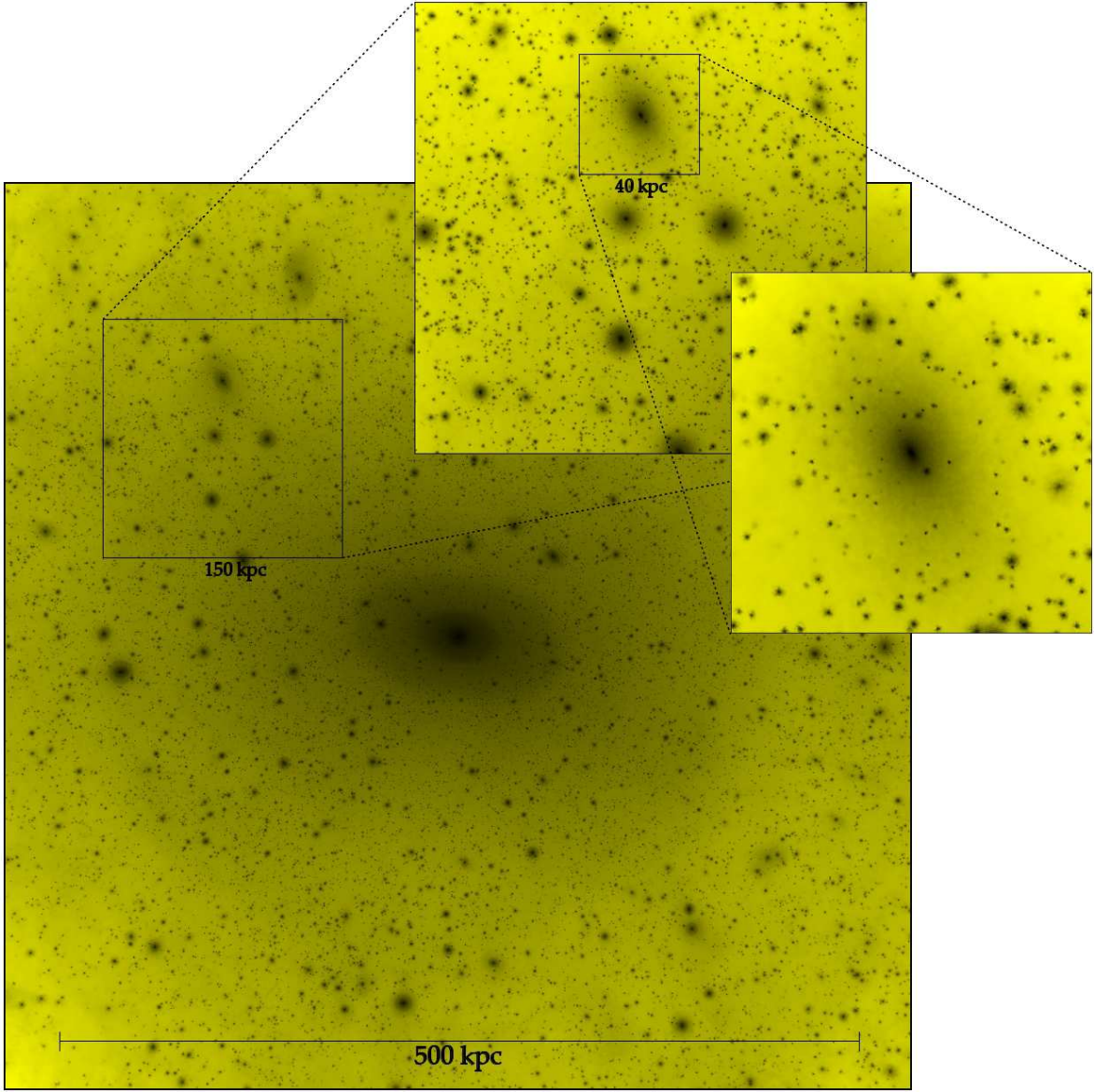}
\caption{A projection of density squared in a 566 kpc $\times$ 566 kpc
  $\times$ 566 kpc region centered on the host halo at $z=0$. The
  insets (150 kpc and 40 kpc deep, respectively) show zooms into a
  region harboring a subhalo exhibiting abundant sub-substructure.}
\label{fig:poster}
\end{center}
\end{figure}

The main scientific driver for our \VL\ simulation is to better
characterize the dark matter subhalo population of a Milky-Way-scale
halo.  A visual impression of the enormous abundance of subhalos is
given by Figure~\ref{fig:poster}, which shows a density squared
projection of the central (566 kpc)$^3$ cube. Projecting density
squared enhances the contrast between individual subhalos and the
smooth background. The two insets show progressive zooms into a region
of interest. The deepest zoom is a projection of a (40 kpc)$^3$ cube
centered on subhalo with $\Vmax=21.4$ km s$^{-1}$\footnote[1]{$\Vmax$
  is the peak of the circular velocity curve ($v_{\rm
    circ}=\sqrt{GM(<r)/r}$) and is used as a proxy for mass.} and a
tidal radius of 46.8 kpc. The high resolution of \VL\ is able to
resolve the second level of the subhalo hierarchy -- subhalos of a
subhalo.

The overall picture is reminiscent of the ``turtles all the way down''
image. In fact, in our simulation the clumpiness extends over six
orders of magnitude in mass down to our subhalo resolution limit of
$\sim 10^5 \msun$. The theoretical expectation is that this clumpiness
continues even further, all the way down to the cut-off in the density
fluctuation power spectrum set by the finite velocity dispersion (or
temperature) of the cold dark matter particle \cite{green2005,
  profumo2006}, some 10 to 15 orders of magnitude below the scales
that we currently resolve. An important caveat here is that our
simulation does not include the effects of baryons. Interactions with
stars or giant molecular clouds in the disk of the Milky Way could
destroy some fraction of the subhalos passing through the center. Such
baryonic affects are unlikely to affect the subhalo population as
whole, however, since most of the subhalos are not on orbits that
carry them through the central disk.

\subsection{The dark matter subhalo population}
\begin{figure}
\begin{center}
\includegraphics[width=\textwidth]{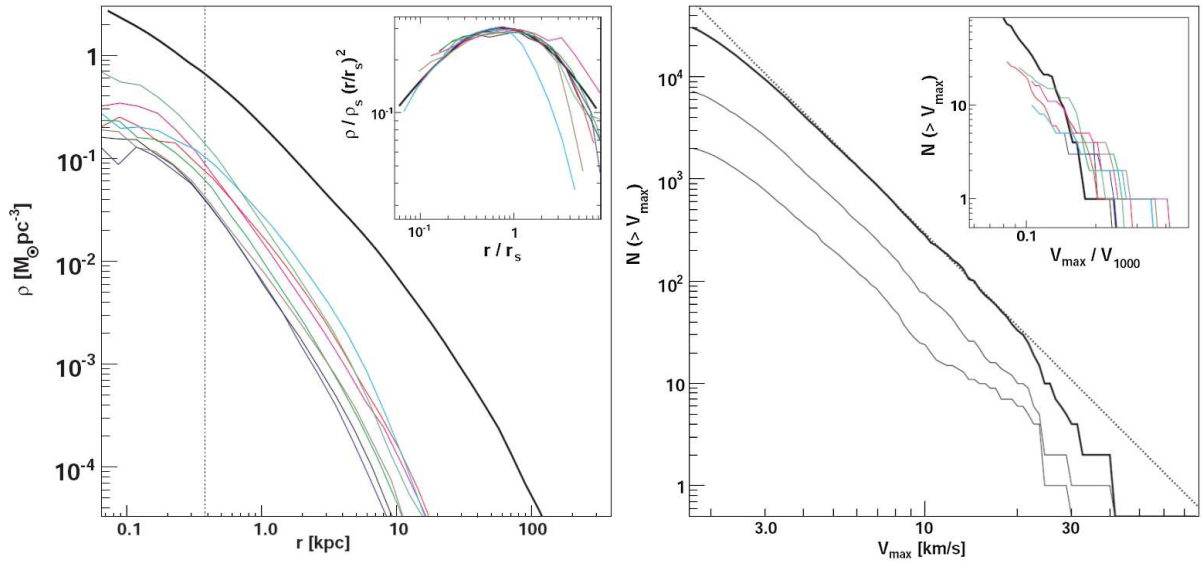}
\caption{Properties of the subhalo population in \VL, from Diemand et
  al. (2008) \cite{diemand2008}. \textit{Left panel:} Density profiles
  of the host halo (thick black line) and of eight large subhalos
  (thin colored lines). The profiles remain cuspy down the estimated
  convergence radius of 380 pc (vertical dotted line). The rescaled
  profiles in the inset demonstrate that i) subhalos are quite
  self-similar in the inner regions and ii) many subhalos are tidally
  truncated in the outer region. \textit{Right panel:} The number of
  subhalos above $\Vmax$ within $\rtwo=402$ kpc (thick line) and
  within 100 and 50 kpc of the halo center (thin lines). $\Vmax$ is
  the peak of the subhalo circular velocity $v_{\rm
    circ}=\sqrt{GM(<r)/r}$, and serves as a proxy for subhalo
  mass. The distribution is well fit by a single power-law, $N(>\Vmax)
  = 0.036 (\Vmax/V_{\rm max, host})^{-3}$ (dotted line). Below $\Vmax
  \simeq 3.5$ km s$^{-1}$ the subhalo abundance is artificially
  reduced due to numerical limitations. The inset shows the abundance
  of sub-subhalos in eight large subhalos, and is quite similar to the
  scaled down version of the host halo.}
\label{fig:subhalos}
\end{center}
\end{figure}

The density profiles of isolated dark matter halos are known to follow
a common radial dependence, falling off as $r^{-3}$ in the outer
regions, becoming shallower towards the center, and finally
approaching a central cusp of $r^{-\gamma}$, with the exact value of
$\gamma$ being somewhat uncertain, but less than 1.5
\cite{navarro1996, moore1999, diemand2004}. The very large number of
particles in the \VL\ simulation and the extremely accurate integration
of their orbits has allowed a precise determination of the density
profile within the inner kpc of the Galactic halo and its satellite
halos, see the left panel of Figure~\ref{fig:subhalos}. The host halo
density profile is best fit with a central cusp of slope
$\gamma=1.24$, and we find that the inner profiles of subhalos are
also consistent with cusps, with slopes scattered around
$\gamma=1.2$. At larger radii the subhalo density profiles generally
fall off faster than the host halo (see inset). The explanation is
simply that subhalo density profiles are modified by tidal mass loss,
which strips material from the outside in, but does not affect the
inner cusp structure \cite{kazantzidis2004, diemand2007}.

Overall \VL\ predicts a remarkably self-similar pattern of dark matter
clustering properties. The right panel of Figure~\ref{fig:subhalos}
shows that subhalos appear to be scaled down versions of their host
halo not just in the inner mass distribution, but also in terms of
their relative substructure abundance. The overall subhalo abundance
as a function of mass, as measured by the cumulative $\Vmax$ function,
is well described by a single power law
\be
N(>\Vmax) \propto (\Vmax/V_{\rm max, host})^{-3},
\ee
independently of distance, within 50, 100, or 402 kpc of the
center. Furthermore, for the first time it has become possible to
determine the abundance of subhalos at the second level of the
substructure hierarchy, and the sub-subhalo $\Vmax$-functions (see
inset) appear to be quite similar to the scaled down version of the
host halo's subhalo distribution.

\subsection{The ``missing satellites problem''}

It has been known for many years that our host galaxy, the Milky Way,
is orbited by a number of so-called dwarf galaxy satellites. The
original census of about 11 dwarfs \cite{mateo1998} has recently
doubled in size by the discovery of $\sim 12$ additional ultra-faint
dwarfs in the Sloan Digital Sky Survey (SDSS) \cite{willman2005,
  zucker2006a, zucker2006b, belokurov2006, belokurov2007,
  grillmair2006, sakamoto2006, walsh2007}. Many, if not all, of these
systems are strongly dark matter dominated in their central regions
\cite{strigari2007, simon2007}, and are thus thought to be the
luminous counterparts to the massive end of the dark matter subhalo
population. The apparent discrepancy between the number of observed
dwarf galaxies (around 50, after correcting for the incomplete sky
coverage of the SDSS) and the vastly larger number of dark matter
subhalos predicted by numerical simulations ($\sim$ 50,000 in \VL) is
well known as the ``missing satellites problem''.

The solution to this problem could lie in a modification of the
standard cold dark matter paradigm of structure formation. For
example, \textit{warm} dark matter, with its higher intrinsic velocity
dispersion, would lead to a suppression of power at small scales and
greatly reduce the number of low mass subhalos seen in CDM
simulations. On the other hand, the standard CDM picture might be
fine, if complicated ``gastrophysics'', such as radiative or thermal
feedback from young galaxies, quasars, or supernovae, prevents star
formation in the vast majority of dark matter subhalos. In this case
the Galactic halo would be populated by an enormous number of
invisible dark matter clumps. The resolution of the ``missing
satellites problem'' depends on both progress from the observational
side: deeper surveys, more and improved measurements of stellar
kinematics in the centers of dwarfs, etc.; and from the theoretical
side: better characterization of the properties of dark matter
subhalos from simulations, a better understanding of the efficacy of
feedback in preventing star formation, etc. Our \VL\ simulations are
one step in this direction \cite{madau2008}.

\subsection{Indirect detection of dark matter substructure?}

An intriguing, if somewhat speculative, possibility for learning more
about the dark matter subhalos of our Milky Way arises if the dark
matter particle is a WIMP, a weakly interacting massive particle, like
the neutralino in the supersymmetric extension to the standard model
of particle physics. In that case the dark matter particle would be
its own anti-particle, and in regions of sufficiently high density
colliding dark matter particles would pair-annihilate. The energetic
standard model particles that are produced in such annihilations would
decay in a particle cascade, resulting in a potentially observable
signal of electron-positron pairs, neutrinos, and gamma-rays. Space
based experiments (like PAMELA and the upcoming AMS-02,
\cite{antiparticles}) could detect anti-particles produced in dark
matter annihilations within about one kpc, and huge neutrino
telescopes built deep into Antarctic ice (e.g. AMANDA-II and its
successor IceCube \cite{neutrinos}) or in the Mediterranean sea
(ANTARES \cite{neutrinos}) have also been pursuing such an indirect
detection. The strength of such a signal depends on the local dark
matter density in the solar neighborhood, and can be boosted by the
clumpiness of the dark matter distribution. Extrapolating from the
measured subhalo abundance in our simulation, we have estimated a
local boost (within 1 kpc) of about 40\% \cite{diemand2008}.

\begin{figure}
\begin{center}
\includegraphics[width=\textwidth]{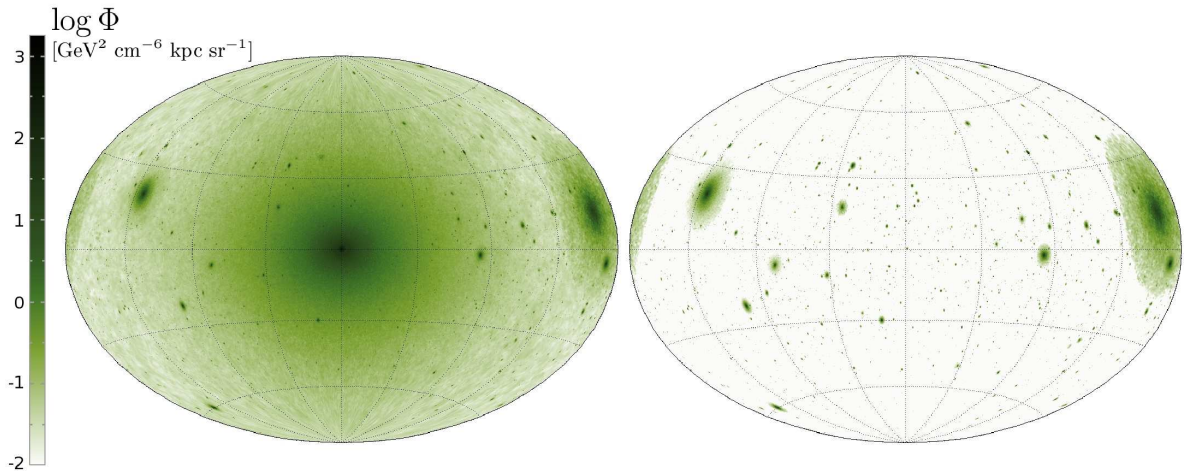}
\caption{Allsky maps (in a Hammer-Aitoff projection) of the
  annihilation signal \mbox{($\propto \int_{\rm los} \rho^2 c^4 \;
    d\ell$)} from \VL, for an observer 8 kpc from the halo
  center. \textit{Left panel:} the total signal from all particles
  within $\rtwo$. \textit{Right panel:} The signal from only the
  subhalo particles. From Kuhlen et al. (2008) \cite{kuhlen2008}.}
\label{fig:allsky}
\end{center}
\end{figure}

The hope for the gamma-ray signal rests on ground-based atmospheric
Cerenkov telescopes (like H.E.S.S., VERITAS, MAGIC, and STACEE,
\cite{acts}) and on NASA's recently launched Gamma-ray Large Area
Space Telescope (GLAST, \cite{glast}). We have used the
\VL\ simulation to make quantitative predictions for the possible dark
matter annihilation signal that GLAST may detect \cite{kuhlen2008}. By
placing fiducial observers at 8 kpc (the distance from the Sun to the
Galactic center) from the host halo center and summing up the
annihilation fluxes from all particles in our simulation, we produced
allsky maps of the gamma-ray flux from dark matter annihilations
within the Galaxy. An example of such an allsky map is shown in
Figure~\ref{fig:allsky}. Converting these flux maps into expected
number counts of gamma-ray photons requires assumptions about the
nature of the dark matter particle, namely its mass, its annihilation
cross section, and the gamma-ray spectrum resulting from a single
annihilation event. Particle physics theory today does not uniquely
predict these quantities, and so we have sampled a region of the
theoretically motivated parameter space. Using detailed models of the
known astrophysical gamma-ray backgrounds that the dark matter
annihilation signal from Galactic substructure has to contend with, we
determined the expected detection significance for every subhalo in
our simulation. Although the vast majority of the subhalos are too
faint to be detectable, we found that for reasonable choices of the
particle physics parameters, GLAST should be able to detect a handful
of individual subhalos at more than five sigma significance.

\section{Summary}
The \VL\ simulation is a one billion particle simulation of the
formation and evolution of a Milky Way scale dark matter halo in an
expanding cosmological volume. At a cost of about 1 million cpu hours
on ORNL's Cray XT3 supercomputer \textit{Jaguar}, it provides an
unprecedented view of the subhalo population in a dark matter halo
like the one harboring our own Milky Way.

\pkdgrav, the code we used for our simulation, is an example of a
hierarchical tree code that uses a multipole expansion of the
gravitational potential to solve the N-body problem in a fast and
accurate manner. In \VL\ we employed for the first time a new
dynamical time step criterion that ensures a highly accurate orbit
integration even for particles in very dense environments, while at
the same time allowing longer time steps for particles in the outer
regions of the halo. This allowed us to follow the central density
profile and resolve the subhalo population closer to the host halo
center than ever before.

We find that the dark matter clustering properties are remarkably
self-similar: isolated halos and subhalos contain about the same
relative amount of substructure and both have cuspy inner density
profiles. The resolution of the ``missing satellites problem'', a well
known discrepancy between the number of luminous dwarf galaxies
orbiting our Galaxy and the predicted number of dark matter subhalos,
will require either a modification of the standard cold dark matter
paradigm of structure formation or baryonic physics that suppresses
star formation in all but a small fraction of subhalos. In the future
it may be possible to indirectly detect dark matter substructure,
through the measurement of electron-positron pairs, neutrinos, or
gamma-rays produced in pair-annihilations of dark matter particles in
high density regions such as the centers of subhalos.


\section*{References}

\begin{thebibliography}{9}

\bibitem{diemand2005}
Diemand J., Moore B., Stadel J., 2005, \nat, 433, 389

\bibitem{springel2005}
Springel, V., et al.\ 2005, \nat, 435, 629

\bibitem{moore1998}
Moore B., Governato F., Quinn T., Stadel J., Lake G., 1998, \apjl, 499, L5

\bibitem{spergel2007}
Spergel, D.~N., et al.\ 2007, \apjs, 170, 377

\bibitem{peebles1980}
Peebles, P.~J.~E. 1980, The large-scale structure of the universe, Princeton University Press, 1980

\bibitem{stadel2001}
Stadel, J.~G. 2001, Ph.D.~Thesis, Univ. of Washington

\bibitem{anderson1996} Anderson, R. 1996, Tree Data Structures for
  {N}-Body Simulation, in FOCS: IEEE Symposium on Foundations of
  Computer Science, \texttt{http://citeseer.ist.psu.edu/177236.html}

\bibitem{dehnen2001}
Dehnen, W.\ 2001, \mnras, 324, 273


\bibitem{power2003}
Power, C., Navarro, J.~F., Jenkins, A., Frenk, C.~S., White, S.~D.~M., Springel, V., Stadel, J., \& Quinn, T. 2003, \mnras, 338, 14

\bibitem{zemp2007}
Zemp, M., Stadel, J., Moore, B., \& Carollo, C.~M.\ 2007, \mnras, 376, 273 

\bibitem{bertschinger2001}
Bertschinger, E. 2001, \apjs, 137, 1

\bibitem{green2005}
Green, A.~M., Hofmann, S., \& Schwarz, D.~J.\ 2005, JCAP, 8, 3

\bibitem{profumo2006}
Profumo, S., Sigurdson, K., \& Kamionkowski, M.\ 2006, Physical Review Letters, 97, 031301 

\bibitem{navarro1996}
Navarro, J.~F., Frenk, C.~S., \& White, S.~D.~M.\ 1996, \apj, 462, 563

\bibitem{moore1999}
Moore, B., Ghigna, S., Governato, F., Lake, G., Quinn, T., Stadel, J., \& Tozzi, P.\ 1999, \apjl, 524, L19

\bibitem{diemand2004}
Diemand, J., Moore, B., \& Stadel, J.\ 2004, \mnras, 353, 624

\bibitem{kazantzidis2004}
Kazantzidis, S., Mayer, L., Mastropietro, C., Diemand, J., Stadel, J., \& Moore, B.\ 2004, \apj, 608, 663

\bibitem{diemand2007}
Diemand, J., Kuhlen, M., \& Madau, P.\ 2007, \apj, 667, 859

\bibitem{mateo1998}
Mateo, M.~L.\ 1998, \araa, 36, 435

\bibitem{willman2005}
Willman, B., et al.\ 2005a, \apjl, 626, L85

\bibitem{zucker2006a}
Zucker, D.~B., et al.\  2006a, \apjl, 643, L103 

\bibitem{zucker2006b}
Zucker, D.~B., et al.\ 2006b, \apjl, 650, L41 

\bibitem{belokurov2006}
Belokurov, V., et al.\ 2006b, \apjl, 647, L111 

\bibitem{belokurov2007}
Belokurov, V., et al.\ 2007b, \apj, 654, 897

\bibitem{grillmair2006}
Grillmair, C.~J.\ 2006, \apjl, 645, L37

\bibitem{sakamoto2006}
Sakamoto, T., \& Hasegawa, T.\ 2006, \apjl, 653, L29

\bibitem{walsh2007}
Walsh, S.~M., Jerjen, H., \& Willman, B.\ 2007, \apjl, 662, L83

\bibitem{strigari2007}
Strigari, L.~E., Bullock, J.~S., Kaplinghat, M., Diemand, J., Kuhlen, M., \& Madau, P.\ 2007, \apj, 669, 676
\
\bibitem{simon2007}
Simon, J.~D., \& Geha, M.\ 2007, \apj, 670, 313 

\bibitem{madau2008}
Madau, P., Diemand, J., \& Kuhlen, M.\ 2008, \apj, 679, 1260

\bibitem{antiparticles}
\begin{minipage}[t]{6in}
\texttt{http://pamela.roma2.infn.it/} \\
\texttt{http://ams.cern.ch/}
\end{minipage}

\bibitem{neutrinos}
\begin{minipage}[t]{6in}
\texttt{http://amanda.uci.edu/} \\
\texttt{http://antares.in2p3.fr/} \\
\texttt{http://icecube.wisc.edu/}
\end{minipage}

\bibitem{diemand2008}
Diemand, J., Kuhlen, M., Madau, P., Zemp, M., Moore, B., Potter, D., \& Stadel, J.\ 2008, Nature, 454, 735

\bibitem{acts}
\begin{minipage}[t]{6in}
\texttt{http://www.mpi-hd.mpg.de/hfm/HESS/HESS.html} \\
\texttt{http://veritas.sao.arizona.edu/} \\
\texttt{http://wwwmagic.mppmu.mpg.de/} \\
\texttt{http://www.astro.ucla.edu/~stacee/}
\end{minipage}

\bibitem{glast}
\texttt{http://glast.gsfc.nasa.gov/}

\bibitem{kuhlen2008}
Kuhlen, M., Diemand, J., \& Madau, P.\ 2008, ApJ, 686, 262

\end{thebibliography}
\end{document}